# Manifestation of transient effects in fission induced by relativistic heavy-ion collisions[a]

B. Jurado[1][b], C. Schmitt[1], K.-H. Schmidt[1], J. Benlliure[2], A. R. Junghans[3]

[1] GSI, Planckstr.1, 64291 Darmstadt, Germany
[2] Univ. de Santiago de Compostela, 15706 S. de Compostela, Spain
[3] Forschungszentrum Rossendorf, Postfach 510119, 01314 Dresden, Germany

**Abstract:** We examine the manifestation of transient effects in fission by analysing experimental data where fission is induced by peripheral heavy-ion collisions at relativistic energies. Available total nuclear fission cross sections of $^{238}$U at 1·$A$ GeV on gold and uranium targets are compared with a nuclear-reaction code, where transient effects in fission are modelled using different approximations to the numerical time-dependent fission-decay width: a new analytical description based on the solution of the Fokker-Planck equation and two widely used but less realistic descriptions, a step function and an exponential-like function. The experimental data are only reproduced when transient effects are considered. The deduced value of the dissipation strength $\beta$ depends strongly on the approximation applied for the time-dependent fission-decay width and is estimated to be of the order of $2 \cdot 10^{21}$ s$^{-1}$. A careful analysis sheds severe doubts on the use of the exponential-like in-growth function largely used in the past. Finally, we discuss which should be the characteristics of experimental observables to be most sensitive to transient effects in fission.

## 1. Introduction

As a typical example of a large-scale motion, nuclear fission stands for one of the most promising tools to study dissipation. In the frame of transport theories, the dynamical evolution of the system along its path to fission is described by the *time-dependent* Langevin [1, 2] equation or by the Fokker-Planck equation of motion [3]. The numerical solution of the latter brings in the concept of transient effects, i.e. the fission decay width $\Gamma_f(t)$ explicitly depends on time. More specifically, the fission-decay width $\Gamma_f(t)$ is inhibited at the earliest times and then it increases continuously until it reaches its asymptotic value. Thus, at the beginning of the process, during a delay of the order of the so-called transient time $\tau_{trans}$ [4], the fission-decay width differs from its asymptotic value $\Gamma_f^K$, originally derived by Kramers [5] by solving the *stationary* FPE.

Due to the high computing time required by the Langevin and Fokker-Planck approaches, the interpretation of experimental data is often performed by comparison with evaporation codes. There, the evolution of the nucleus is governed by the competition between fission and particle emission, the probability of each de-excitation channel being weighted by the corresponding decay width. In most evaporation codes, the fission width does not depend on time and is represented by Kramers stationary value $\Gamma_f^K$. Such a simplification may be prejudicial to a reliable description of the evolution of the system. Indeed, the increased particle emission probability at the earliest times, caused by transient effects, influences the further evolution of the nucleus, and in particular its fission probability in a considerable way. A rigorous method that correctly accounts for the time-dependence of the fission decay width would require solving the equation of motion at each evaporation step. In many cases, this is again unconceivable in praxis due to the computational effort required. Nevertheless, any analytical approximation of $\Gamma_f(t)$ which well reproduces the exact numerical solution could easily be implemented in the evaporation code and make the latter equivalent to a dynamical Langevin or Fokker-Planck approach. In our previous paper [6] we carefully investigated the main features of the relaxation process of the probability distribution in deformation and its conjugate momentum, and accordingly derived a new analytical approximation for the fission-decay width $\Gamma_f(t)$ [6, 7]. In addition, we define which are the characteristics of experimental observables needed to be most sensitive to transient effects in fission. For this purpose, we meticulously compare different model calculations including or excluding transient effects.

---

[a] This work forms part of the PhD thesis of B. Jurado
[b] Corresponding author: B. Jurado, GANIL, Boulevard Henry Becquerel, B.P. 5027, 14076 Caen CEDEX 5, France, e-mail: jurado@ganil.fr (present address)

In addition, at high excitation energy not only the relaxation process towards equilibrium has to be considered but thermal instabilities [8] play a non-negligible role in suppressing fission at high excitation energies. Also this issue will be tackled below.

## 2. Experimental approach

During the last decade a large amount of experimental work to study dissipation has been carried out. While it has been quite well established that dissipation is rather strong in the large-distortion range [9, 10, 11, 2], the situation is still unclear in the small-deformation regime [12, 13, 14, 15, 16]. One of the reasons for the vivid debates concerning the strength of nuclear dissipation is the large amount of complex side effects entering into the description of the fission mechanism, in particular in fusion-fission experiments, which often induce large angular momenta and deformations. In the present work, we focus on the investigation of nuclear friction at small shape distortion and high excitation energy by means of a reaction mechanism that leads to fission with considerably less side effects. This aim was achieved by applying a projectile-fragmentation reaction, i.e. a very peripheral nuclear collision with relativistic heavy ions, introducing small shape distortions [17] and low angular momenta [18]. Moreover, compared to previous quite interesting attempts to reach this goal by relativistic proton-nucleus collisions in inverse kinematics [19] and the annihilation of antiprotons [20, 21] at the nuclear surface, this approach populates higher excitation energies more strongly, up to the onset of multifragmentation [8]. Such high excitation energies are crucial for the manifestation of transient effects, as we will show.

## 3. Model calculations

### 3.1 Dynamical description of fission by the ABRABLA code

In the present work, the experimental data will be compared with several calculations performed with an extended version of the abrasion-ablation Monte-Carlo code ABRABLA [17, 22]. This code consists of three stages. In the first stage the properties of the nucleus after the fragmentation reaction are calculated according to the geometrical abrasion model. As was determined experimentally in reference [23], an average excitation energy of 27 MeV per nucleon abraded is induced. This value is in agreement with predictions for peripheral collisions based on BUU calculations [24]. The angular momentum deduced is estimated according to reference [18]. The second stage of the code accounts for the simultaneous emission of nucleons and clusters (simultaneous break up) [25] that is assumed to take place due to thermal instabilities when the temperature of the projectile spectator exceeds 5.5 MeV. The break-up stage is assumed to be very fast, and thus the fission degree of freedom is not excited. This simultaneous emission makes the nucleus cool down to a temperature of 5.5 MeV. From this moment on the sequential decay as the third stage sets in. It is described in a time-dependent formulation and treats the deexcitation of the nucleus by the competition between particle evaporation and fission. To account for transient effects, we incorporate in this last part different descriptions of the time-dependent fission-decay width $\Gamma_f(t)$. Two of them correspond to the most widely used in the past:

- a step function

$$\Gamma_f(t) = \begin{cases} 0, t < \tau_{trans} \\ \Gamma_f^k, t \geq \tau_{trans} \end{cases} \quad (1)$$

where the Kramers decay width $\Gamma_f^K$ corresponds simply to the statistical Bohr-and-Wheeler width [26] multiplied by the Kramers factor [5].

- and an exponential in-growth function

$$\Gamma_f(t) = \Gamma_f^k \cdot \{1-\exp(-t/\tau)\} \quad (2)$$

with $\tau = \tau_{trans}/2.3$ have been included avoiding several further approximations applied in previous formulations [27, 28, 29, 30].

Furthermore, we implemented the description of $\Gamma_f(t)$ of [31] based on a realistic analytical approximation to the solution of the FPE introduced in ref. [6]. The initial conditions corresponding to the zero-point motion are taken into account. These are close to what is encountered in heavy-ion collisions at relativistic energies [32]. A detailed description on how these $\Gamma_f(t)$ functions have been implemented can be found in the appendix **A1**. In addition to the three *time-dependent* expressions, we will also perform calculations using the Kramers *time-independent* decay width $\Gamma_f^K$ which, in contrast to the above-mentioned descriptions, does not include any transient effect at all.

Besides the treatment of dissipation effects, the ratio of the level-density parameters $a_f/a_n$ and the fission barriers $B_f$ are the most critical ingredients of the model. The deformation dependence of the level-density parameter has been discussed in references [33, 34, 36, 35]. In our case, the ratio $a_f/a_n$ is calculated considering volume and surface dependencies as proposed in reference [36] according to the expression:

$$a = \alpha_v A + \alpha_s A^{2/3} B_s \qquad (3)$$

where $\alpha_v$ and $\alpha_s$ are the coefficients of the volume and surface components of the single-particle level densities, respectively, with the values $\alpha_v$ =0.073MeV$^{-1}$ and $\alpha_s$ = 0.095 MeV$^{-1}$. $B_s$ is the ratio of the surface of the deformed nucleus related to the corresponding value of a spherical nucleus. Its value is taken from ref. [37]. A recent work of Karpov et al. [35] has shown that equation (44) is well adapted by comparing it to several derivations: in the framework of the liquid-drop model including a Coulomb term [38], with the finite-range liquid-drop model [39] and within the relativistic mean-field theory [40]. The angular-momentum-dependent fission barriers are taken from the finite-range liquid-drop model predictions of Sierk [41]. As demonstrated in ref. [42], a recent experimental determination of the level-density parameter and the fission barriers by K. X. Jing and co-workers [43], based on the measurement of cumulative fission probabilities of neighbouring isotopes, is in very good agreement with the theoretical parameterisations we use.

### 3.2. Total nuclear fission cross sections

In this work, we analyse the total nuclear fission cross sections of $^{238}$U at 1·$A$ GeV on gold and uranium targets which have been determined by Rubehn et al. [44]. The values are listed in table 1.

In addition to the different shapes of the time-dependent fission width and the different values of $\beta$, for part of the calculations presented in table 1 the break-up stage of the code was not included. Though this is unphysical, it serves to distinguish between the effects of dissipation and those of the break-up process on fission at high excitation energies. These two effects are discussed in detail in the next section separately.

Let us consider the calculations performed including the break-up stage between the abrasion and the sequential decay at temperatures larger than 5.5 MeV. As the values of the fourth row of table 1 show, the transition-state model clearly overestimates the experimental cross sections. In fact, the experimental values are only reproduced when dissipative effects are included in the calculation. However, the choice of the in-growth function for $\Gamma_f(t)$ according to equations (1) or (2) has a strong influence on the dissipation coefficient deduced. While the calculation with the step function reproduces the data with a value of $\beta = 2 \cdot 10^{21}$ s$^{-1}$, the same value of $\beta$ with the exponential in-growth function overestimates the cross sections. The reason is that in the latter case fission is already possible with a non-negligible probability at the very beginning of the de-excitation process. To reproduce the data when the exponential-like in-growth function is used, a larger value of the dissipation coefficient $\beta = 4 \cdot 10^{21}$ s$^{-1}$ is required that reduces the asymptotic value of the fission width and enlarges the transient time. As expected, when this value of $\beta$ is used with the step function, the cross sections are underestimated. A similar conclusion can be drawn from the value of $\beta$ needed to reproduce the data when the Kramers fission width is used. As Kramers' picture does not include any transient time, fission is not inhibited at the earliest times at all, and an even larger value of $\beta$ = 6·10$^{21}$ s$^{-1}$ is required. For the analytical approximation of ref. [6], the noticeable suppression of fission at small times leads to the correct experimental cross section for a smaller value of $\beta = 2 \cdot 10^{21}$ s$^{-1}$.

|  | $\sigma_f^{nucl}$ on Au / b | | $\sigma_f^{nucl}$ on U / b | |
|---|---|---|---|---|
| **Experimental Data** | 2.14 ± 0.22 | | 2.19 ± 0.44 | |
| **Calculation** | No break up | Break up | No break up | Break up |
| Transition-State Model | 5.53 | 3.28 | 5.80 | 3.39 |
| Transition-State Model*Kramers $\beta = 2 \cdot 10^{21}$ s$^{-1}$ | 5.22 | 2.87 | 5.46 | 2.96 |
| Transition-State Model*Kramers $\beta = 6 \cdot 10^{21}$ s$^{-1}$ | 4.64 | 2.16 | 4.86 | 2.24 |
| $\Gamma_f(t)$ step $\beta = 2 \cdot 10^{21}$ s$^{-1}$ | 2.15 | 2.04 | 2.20 | 2.03 |
| $\Gamma_f(t)$ step $\beta = 4 \cdot 10^{21}$ s$^{-1}$ | 1.58 | 1.50 | 1.59 | 1.56 |
| $\Gamma_f(t) \sim 1 - \exp(-t/\tau)$ $\beta = 2 \cdot 10^{21}$ s$^{-1}$ | 4.92 | 2.52 | 5.16 | 2.61 |
| $\Gamma_f(t) \sim 1 - \exp(-t/\tau)$ $\beta = 4 \cdot 10^{21}$ s$^{-1}$ | 4.31 | 2.02 | 4.50 | 2.06 |
| $\Gamma_f(t)$ FPE $\beta = 2 \cdot 10^{21}$ s$^{-1}$ | 2.28 | 2.08 | 2.39 | 2.13 |

**Table 1:** Experimental total nuclear fission cross sections of $^{238}$U(1·$A$ GeV) on gold and uranium targets compared with different calculations performed with the code ABRABLA. The experimental cross sections are taken from [44]. Each calculation has been performed twice. In one case, the simultaneous break-up stage is not included in the calculation, so that no limit for the initial temperature of the sequential decay is imposed. In the other case, the break-up model imposes an upper limit of 5.5 MeV to the initial temperature of the fission-evaporation cascade. The calculations listed in the fourth row were performed with the transition-state model [26]. The ones shown in the fifth and sixth row use the transition-state model, corrected by Kramers factor [5] to account for the reduction of the stationary fission-decay width by dissipation. The other calculations were performed with different descriptions for $\Gamma_f(t)$ and different values of $\beta$ (see text).

From our considerations we conclude that the exponential-like in-growth function fails to model the relaxation process of the nucleus. When used to extract the value of the dissipation coefficient from comparisons of experimental fission cross sections with model calculations, the unrealistically early onset of the fission decay has to be compensated by a similarly unrealistically large suppression of fission in the stationary regime, leading to an overestimation of the dissipation strength. Although the step function appears to be a rather crude approximation, it better describes the effects of dissipation on the time-dependence of the fission-decay width: indeed, it also leads to strong suppression of fission during the relaxation phase like the exact solution of the FPE and our analytical approximation. The analysis of the measured fission cross sections constrains the magnitude of the reduced dissipation coefficient to values in the vicinity of critical damping, around $\beta = 2 \cdot 10^{21}$ s$^{-1}$. This is a valuable result, which fixes the parameters of the model calculations we will present in the following sections.

### 3.3. Break up and time dependence of $\Gamma_f(t)$

As mid-peripheral heavy-ion collisions allow to produce nuclei with excitation energies that are far beyond the onset of multi-fragmentation [8], a break-up stage has been introduced in the ABRABLA code [17] as described above. To clearly distinguish between the effect of the break-up stage and the influence of the time-dependent behaviour of the fission width, let us first compare some results obtained with and without the break-up stage for the four fission-width descriptions used in the calculations presented in table 1. The spectra obtained from calculations performed without break up are shown in Figure 1 and those performed with break up in Figure 2. In both cases the excitation energy with which the nucleus crosses the saddle point is displayed as a function of the excitation energy of the prefragment before the sequential decay stage. Each calculation was done for the value of $\beta$ that reproduces the experimental total nuclear fission cross section for the corresponding fission-width approximation $\Gamma_f(t)$, which has been deduced from table 1. This study has been performed for a lead target, for which the total cross section has been investigated in ref. [7] already. Comparing Figures 1 and 2, we see that whatever the expression of $\Gamma_f(t)$, the inclusion of the break-up process leads to a suppression of the fission events with the largest excitation energy at saddle. This feature is obviously directly connected to the treatment of break up in the ABRABLA code for abrasion products above a temperature of 5.5 MeV, corresponding to an initial excitation energy of the order of 800 MeV. Whereas the influence of the break-up stage is rather drastic in the case of the Kramers stationary fission-decay width and the exponential-like fission-decay width, it is quite tiny for the two other descriptions of $\Gamma_f(t)$. This explains that the influence of the break up on the total nuclear fission cross section strongly depends on the description of the fission width as could be seen already in table 1.

The more or less striking effect of the break-up process is connected to the time behaviour of the fission-decay width $\Gamma_f(t)$. The too early onset of fission introduced by the exponential-like in-growth function implies that fission at high excitation energies is hardly suppressed by dissipation and mostly constricted by the break-up mechanism. A similar and even more obvious trend is observed with Kramers stationary width. Among all considered expressions of $\Gamma_f(t)$ this one leads to the highest fission probability at the earliest times, so that the number of systems with very high initial excitation energy is even larger than for the exponential-like in-growth function. On the contrary, the calculations performed with the step function or with the analytical solution of the FPE illustrate that dissipation considerably inhibits the fission decay channel for excitation energies at fission $E^*_{saddle}$ above around 350 MeV, and, therefore, fission does not reach the energy threshold at which break up sets in. In ref. [6] we have shown that for our analytical approximation to the solution of the FPE the fission probability sets in earlier than for the step function, leading to a larger amount of fissioning systems with initial excitation energies $E^*_0$ beyond 800 MeV, as can be seen comparing Figures 1 b) and 1 d). As a result, the break-up stage has a stronger effect in this case than for the step function.

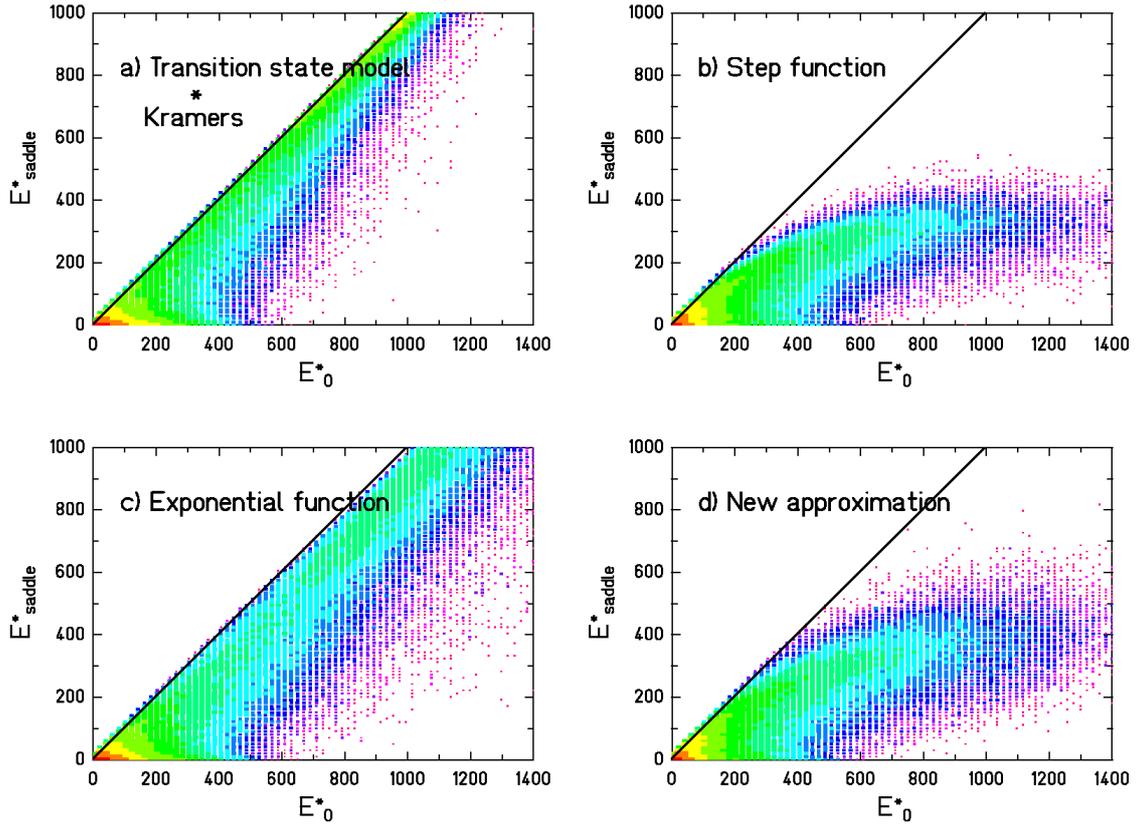

**Figure 1**: Calculations performed with ABRABLA without including the break-up stage. They represent the excitation energy at fission versus the excitation energy of the prefragment after abrasion for the fission events obtained in the reaction of 1·$A$ GeV $^{238}$U on a lead target. a) Calculation performed with the Kramers' fission-decay width and $\beta = 6\times10^{21}$ s$^{-1}$ b) Calculation performed with $\Gamma_f(t)$ as a step function and $\beta = 2\cdot10^{21}$ s$^{-1}$. c) Calculation with $\Gamma_f(t) \propto (1-e^{-t/\tau})$ and $\beta = 4\times10^{21}$ s$^{-1}$. d) Calculation carried out with the analytical approximation of ref. [6] and $\beta = 2\cdot10^{21}$ s$^{-1}$. In all cases we have used $\hbar\omega_g = 1$ MeV. The straight lines represent the upper limit for the excitation energy at fission.

The 45-degree straight lines depicted on the spectra of Figures 1 and Figures 2 correspond to the upper limit for the excitation energy at fission $E^*_{saddle}$. Both, the calculation performed with the step function on Figure 2 b) and the calculation with the analytical solution of the FPE on Figures 2 d), show that this line starts to be depopulated at initial excitation energies $E^*_0$ of approximately 150 MeV. This indicates that from these excitation energies on the transient time is longer than the decay time for particle emission: Therefore, prefragments formed with initial excitation energies higher than around 150 MeV can only fission after cooling down by particle evaporation, since fission is *delayed* by dissipation. In the following calculations we will always include the break up in the decay process at sufficiently high initial excitation energies.

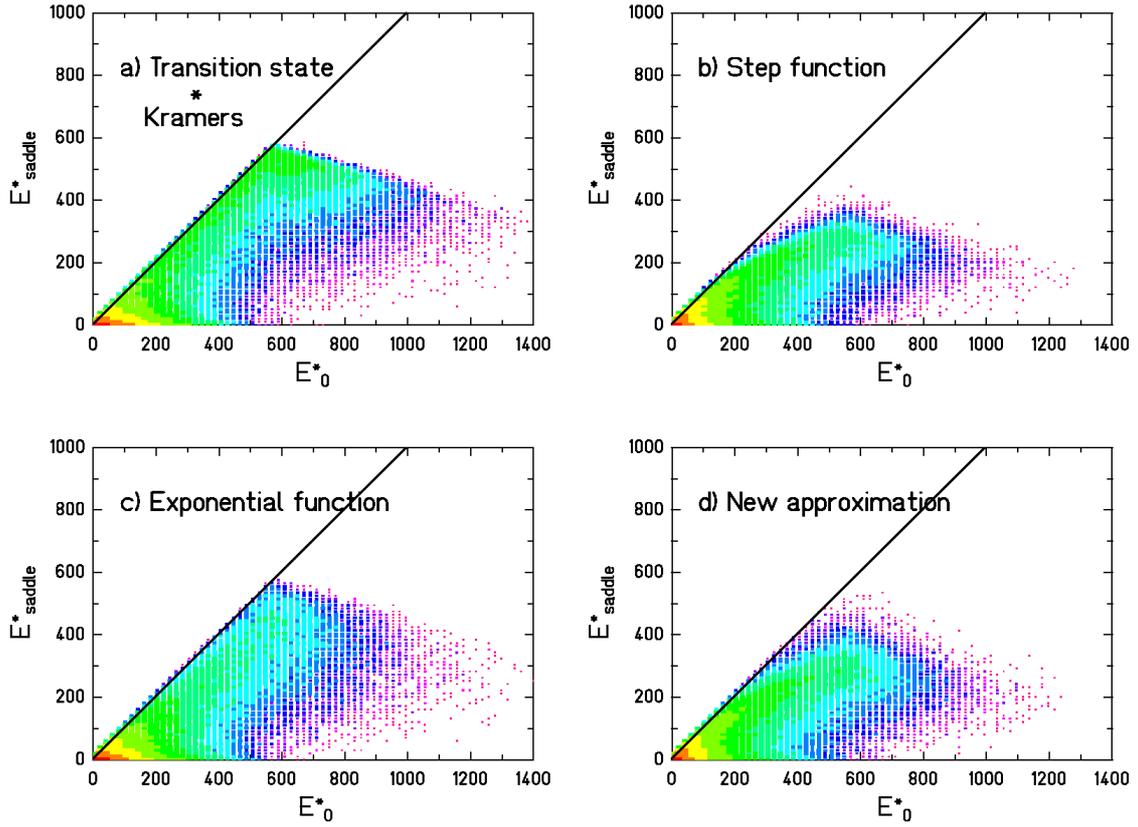

**Figure 2**: The same calculations as in Figure 1 but including the break-up stage in ABRABLA. The cluster plots represent the excitation energy at fission versus the excitation energy of the prefragment before the break-up stage for the fission events obtained in the reaction of 1·$A$ GeV $^{238}$U on a lead target.

*3. 4. Manifestation of transient effects*

The characteristic differences between the four calculations in Figure 2 are directly connected to the differences in the description of transient effects, i.e. in initial behaviour of the in-growth functions used. Furthermore, these differences help to define experimental signatures sensitive to transient effects. Indeed, Figures 2 shows that the calculated excitation energy at saddle depends quite strongly on the shape of the in-growth function $\Gamma_f(t)$. Moreover, our investigation also points out that this dependence appears in a restricted energy range only, namely between $E^*_{saddle}$ = 150 MeV and 350 MeV. Consequently, an experimental observable directly connected to the excitation energy at saddle might be used as a signature for transient effects. In the following paper [45], we will introduce two new signatures, which fulfil this requirement.

## 4. Discussion

We have shown that the relaxation of a system towards equilibrium has a strong influence on the decay process. Consequently, the stationary picture of Kramers cannot be applied to non-equilibrated systems.

Although the step function appears to be a rather crude approximation, it seems to be better suited to describe the effects of dissipation on the fission-decay width $\Gamma_f(t)$ than the exponential-like function. It also leads to a strong suppression of fission at high excitation energies in good agreement with the more realistic description of $\Gamma_f(t)$ given by the analytical approximation to the solution of the FPE proposed in reference [6]. When the step function is used, the experimental total fission cross sections are reproduced with a reduced dissipation coefficient of $\beta=2\cdot 10^{21} s^{-1}$. The same value is obtained for the more realistic description of $\Gamma_f(t)$ based on the analytical approximation. One would expect that this description requires a larger value of $\beta$ to describe the experimental data than the step function, but as explained before, the break-up mechanism suppresses the additional fission events at high excitation energies and leads to very similar fission cross sections.

Our investigation does not allow a direct conclusion on the temperature dependence of the dissipation coefficient $\beta$, but it reveals the difficulty in deducing such an effect. The different distributions of excitation energies at fission found in the calculations shown in Figures 1 and 2 prove that this analysis strongly depends on the in-growth function assumed for $\Gamma_f(t)$. For instance, the inhibition of fission at high excitation energies obtained with the analytical approximation and the step function could be achieved with the exponential-like in-growth function as well by setting very high values of $\beta$ at high temperatures. Therefore, it cannot be excluded that the indications for a strong increase of nuclear viscosity with increasing temperature drawn in several publications [46, 47] is related to the unrealistic exponential-like in-growth function used in the analysis.

The challenge to observe transient effects is increased by the fact that they show up only in a restricted energy range. Indeed, our model calculations have shown that fission is affected by transient effects only for excitation energies at saddle within the interval 150 MeV $< E^*_{saddle} <$ 350 MeV. Above this range fission is almost completely inhibited, independently of the strength of $\beta$. At excitation energies below 150 MeV, the statistical decay times for fission and particle emission are appreciably longer than typical dynamical time scales, making the experimental observables rather insensitive to the transient time. This point may explain why in several experiments performed at rather low excitation energy no transient effects at all were observed [12, 13].

As mentioned in the introduction, nuclear dissipation is expected to depend on temperature and deformation, but the behaviour of these dependences are far from being well established, mainly due to the difficulty to extract clear signatures of the different influences. The necessity for clarifying the role of deformation and temperature is exemplified in reference [48] in which the authors were able to reproduce their data by applying either a temperature-dependent or a deformation-dependent dissipation. Because our results are based on the analysis of fission cross sections, they are only sensitive to dissipation in the small deformation range from the ground state to the saddle point. Better information on dissipation in this restricted deformation regime will help to improve our knowledge on the deformation dependence of the dissipation strength. The value of $\beta = 2\cdot 10^{21}$ $s^{-1}$ obtained from our analysis coincides with the value found in ref. [19] for the reaction Au (800·$A$ MeV) + p. Other work [49, 50, 51] sensitive to the same deformation range is consistent with our conclusions, although often only upper limits for the transient time or the dissipation coefficient could be deduced.

The quantitative value deduced for the dissipation coefficient remains model dependent to a certain degree. Nevertheless, variations of the most critical model parameters by reasonable amount: excitation energy of the prefragments by 30%, freeze-out temperature by 20 % and excitation energy reduction per mass loss in the break-up stage by a factor of two did not modify the value of the deduced dissipation coefficient $\beta$. The most important achievements of the present work, however, are the detailed discussion of an experimental method suited for the study of dissipation at small deformation and the appropriate modelling of dissipation effects in a dynamical nuclear-reaction code.

## 5. Conclusion

Peripheral heavy-ion collisions at relativistic energies are a very suitable tool for investigating dissipation, because highly excited nuclei are produced with initial conditions that exclude many complex side effects. We analysed the nuclear-induced total fission cross sections of $^{238}$U at 1·$A$ GeV on two different heavy targets by means of the Monte-Carlo code ABRABLA. We implemented in this code the two most widely used descriptions for the time dependence of the fission-decay width, a step function and an exponential-like in-growth function, as well as our new analytical approximation. The dissipation strength $\beta$ deduced depends on the shape of the time-dependent fission-decay width $\Gamma_f(t)$. Consequently, in order to interpret any result on the magnitude of $\beta$, the description used for the time dependence of the fission width must be specified. A careful analysis of the exponential-like in-growth function has shown that it fails in reproducing the total suppression of fission at the beginning of the de-excitation process that is expected theoretically. On the contrary, the step-function approximation and the proposed analytical parameterisation show a similar behaviour. This indicates that the suppression of the fission decay width during the initial time lapse is needed to account for dissipation effects in a proper way. The present analysis clearly demonstrates that transient effects in fission are accessible to experimental investigations. Using the analytical approximation of the Fokker-Planck solution, the experimental data are reproduced when $\beta = 2 \cdot 10^{21}$ s$^{-1}$.

**Acknowledgement**

We acknowledge valuable discussions with Hans Feldmeier, Anatoly V. Ignatyuk, and David Boilley. This work has been supported by the European Union in the frame of the HINDAS project under contract FIKW-CT-2000-0031 and by the Spanish MCyT under contract FPA2002-04181-C04-01. One of us (C. S.) is thankful for the financing of a one-year stay at GSI by a Humboldt fellowship. The work profited from a collaboration meeting on "Fission at finite thermal excitations" in April 2002, sponsored by the ECT* ("STATE" contract).

# APPENDIX

## A1: Treatment of fission as a dissipative process in ABRABLA

In this appendix we document in detail how dissipation is treated in the description of the deexcitation of a heavy excited nucleus within the Monte-Carlo code ABRABLA [17, 22]. The main evaporation channels we consider are neutron, proton and alpha-particle emission. Each of these decay modes is represented by a partial decay width $\Gamma_v$. In the following, we distinguish only between two decay channels, particle evaporation, which groups neutron, proton and alpha-particle emission, and fission. The particle-decay mode is represented by the width

$$\Gamma_p = \sum_v \Gamma_v \qquad (A1.1)$$

In ABRABLA the particle-decay widths do not explicitly depend on time. On the contrary, dissipation effects lead to a time-dependent fission width $\Gamma_f(t)$. Starting at $t = 0$ from the initial conditions given by the entrance channel, e.g. the abrasion process, and later by the conditions after each de-excitation step, the functional form of $\Gamma_f(t)$ in the full range $0 \leq t \leq \infty$ is given by the solution of the FPE or eventually by some analytical approximation. That means that the fission-decay width $\Gamma_f(t)$ and the particle decay widths $\Gamma_v$ are newly calculated after each de-excitation step for the corresponding nucleus and its actual excitation energy. In this process, one has to consider that, while particles are emitted, the probability distribution in deformation becomes broader, and at each step of the de-excitation cascade the distribution in deformation "remembers" its current width. This effect is taken into account in an approximate way by evaluating the fission-decay width $\Gamma_f(t)$ in every step with a time offset at step $n$ that is given by the sum of the decay times in the previous steps:

$$t_{sum}^n = \sum_i^{n-1} t_i \qquad (A1.2)$$

The decay time $t_i$ of the decay step $i$ is taken at random from the total decay-time distribution including all decay modes.

In general, as already mentioned above, there exist three methods to model the decay of a heavy excited nucleus in a dynamical way, on one side we have the methods based on the solution of Langevin or the FP equations in small time steps, and on the other side the evaporation codes. In order to understand the method followed in the ABRABLA code, it is useful to clarify the differences and the correspondences of these three different dynamical approaches. The sequence of the computation steps of each of them is summarized in table **A1.1**. Although the computation steps differ, the third option corresponds directly to the second one, since in both cases the probability distribution of the nucleus in deformation space is considered instead of following individual trajectories in the first option. In ABRABLA we incorporated the two approximations for $\Gamma_f(t)$ represented by equations (1) and (2) and the more realistic case given by the equations introduced in ref. [7]. As explained below, for the step function and the exponential-like approximation the third option of table A1.1 is applied throughout the calculation. In contrast, for the analytical approximation the second option is applied before reaching the stationary state, although the FPE is replaced by the analytical approximation, and the third option is applied for later times. In case of deformation-dependent particle-decay widths, which are actually not considered, they would have to be averaged over the actual deformation distribution in the corresponding time steps.

Also the variation of the available intrinsic excitation energy as a function of deformation [52] could be taken into account in an averaged way. Both effects do not have a strong influence on the results [52].

| Stochastic approaches | | Evaporation code |
|---|---|---|
| Langevin equation with allowance for evaporation | Fokker-Planck equation with allowance for evaporation | Evaporation with time-dependent fission width |
| 1. Calculate next position of individual trajectory in a small time step by solving the discretised Langevin equation.<br>2. Determine particle evaporation at random. The probability is small in one time step.<br>3. Back to 1. | 1. Calculate next distribution in deformation by solving the Fokker-Planck equation in a small time step.<br>2. Determine particle evaporation or fission at random. The probability is small in one time step.<br>3. Back to 1. | 1. Calculate complete partial and total decay-time distributions for ($0 \leq t \leq \infty$) (specific to $Z$, $A$, $E^*$); $dI_f/dt$ from the FPE or from an analytical approximation; particle decay-time distributions are exponentials.<br>2. Determine next decay time at random from the total decay-time distribution $dI_{tot}(t - t^n_{sum})/dt$.<br>3. Determine type of decay at random according to the values of the partial decay-time distributions $dI_i/dt$ at the chosen decay time.<br>4. Back to 1. |

**Table A1.1:** The main computation steps in three equivalent dynamical approaches to describe the deexcitation process of an excited nucleus.

For describing the shape of $\Gamma_f(t)$ we incorporated in the same code the two approximations represented by equations (1) and (2) and the more realistic case given by equations introduced in ref. [7]. In all cases the stationary value of the fission width is

$$\Gamma_f^{K,n} = \Gamma_f^{BW,n} \cdot K \tag{A1.3}$$

where $\Gamma_f^{BW,n}$ is the fission width given by the transition-state model [26] at the step $n$ and $K$ is the Kramers factor. Detailed formulations of the three cases are given below. We would like to stress that in all three cases the treatment in ABRABLA exactly corresponds to the schemes sketched in the second, respectively the third column of table 1.1, although the formulations are written below in an analytical and more compact form, if this is possible.

## A1.1. Description of $\Gamma_f(t)$ by a step function

The fission width can be described by a step function that sets in with the stationary value given by equation (A1.3) at the transient time $\tau_{trans}$. Let $I_0$ be the number of nuclei available at the beginning of a certain deexcitation step $n$. For the time interval $t^n_{sum} < t < \tau_{trans}$, particle emission is the only deexcitation channel available, and the number of nuclei that decay in that step $n$ is

$$I_0 \cdot (1 - \exp(-(\tau_{trans} - t^n_{sum})/\tau^n_p)) \tag{A1.4}$$

The upper index $n$ indicates that these quantities differ from step to step. For $t \geq \tau_{trans}$, fission is also possible, and the additional number of nuclei that decay by particle emission at the same step is

$$I_0 \cdot \exp(-(\tau_{trans} - t^n_{sum})/\tau^n_p) \cdot \frac{\Gamma^n_p}{\Gamma^n_p + \Gamma^{K,n}_f} \tag{A1.5}$$

where the quantity $I_0 \cdot \exp(-(\tau_{trans} - t^n_{sum})/\tau^n_p)$ represents the nuclei that survived particle emission before $\tau_{trans}$. Similarly, the number of nuclei that fission after $\tau_{trans}$ is

$$I_0 \cdot \exp(-(\tau_{trans} - t^n_{sum})/\tau^n_p) \cdot \frac{\Gamma^{K,n}_f}{\Gamma^n_p + \Gamma^{K,n}_f} \tag{A1.6}$$

The total probability for particle evaporation at a step $n$ can be obtained by normalizing the total number of decays to the initial number of nuclei $I_0$ leading to the expression

$$P^n_p = (1 - \exp(-(\tau_{trans} - t^n_{sum})/\tau^n_p)) + \exp(-(\tau_{trans} - t^n_{sum})/\tau^n_p) \cdot \frac{\Gamma^n_p}{\Gamma^n_p + \Gamma^{K,n}_f} \tag{A1.7}$$

and the total probability for fission is

$$P^n_f = \exp(-(\tau_{trans} - t^n_{sum})/\tau^n_p) \cdot \frac{\Gamma^{K,n}_f}{\Gamma^n_p + \Gamma^{K,n}_f} \tag{A1.8}$$

This formulation is similar but more consistent than the approach proposed previously in [53] and [36].

## A1.2. Description of $\Gamma_f(t)$ by an exponential in-growth function

Another possibility to describe the time dependence of the fission width $\Gamma_f(t)$ is an exponential in-growth function. In this case, the procedure is different than in the previous one because $\Gamma_f(t)$ increases continuously with time.
The decay rate at a step $n$ can be written as:

$$\left(\frac{dI}{dt}\right)_n = -\left[\frac{\Gamma_p^n}{\hbar} + \frac{\Gamma_f^n(t)}{\hbar}\right] \cdot I^n(t) = -\left[\frac{\Gamma_p^n}{\hbar} + \frac{\Gamma_f^{K,n}}{\hbar}(1-\exp(-\frac{t+t_{sum}^n}{\tau}))\right] \cdot I^n(t) \quad (A1.9)$$

By integrating equation (A1.9) we obtain an analytical expression for $I^n(t)$

$$I^n(t) = I_0 \cdot \exp(-\frac{\Gamma_p^n + \Gamma_f^{K,n}}{\hbar}t) \cdot \left[\exp(-\frac{\Gamma_f^{K,n} \cdot \tau}{\hbar} \cdot \exp(-\frac{t+t_{sum}^n}{\tau}))\right] \quad (A1.10)$$

Substituting equation (A1.10) in (A1.9) we obtain an analytical expression for the decay rate $(dI(t)/dt)_n$ of the nucleus that includes all possible particle decay channels and the already mentioned time dependence of $\Gamma_f$. After evaluating $(dI(t)/dt)_n$ for the corresponding step, we sample from $(dI(t)/dt)_n$ the decay time and finally calculate the value of $\Gamma_f(t)$ at that decay time. The decay channel is then determined by a Monte-Carlo selection with the weights $\Gamma_f(t)/\Gamma_{total}$ for fission and $\Gamma_v/\Gamma_{total}$ for the emission of the particle $v$.

**A1.3. Description of $\Gamma_f(t)$ by the analytical solution of the FPE for a parabolic nuclear potential**

As shown in ref. [6], our new analytical solution of the FPE when the nuclear potential is approximated by a parabola gives a much more realistic description of the time dependence of the fission rate $\lambda_f(t)$ than the step function and the exponential-like in-growth function. Actually, the numerical method [54] we use to include this description of $\Gamma_f(t)$ in the evaporation code is applicable for any function representing $\Gamma_f(t)$. In order to account for the variation of $\Gamma_f(t)$ at small times in the order of the transient time $\tau_{trans}$, we first divide the time interval $0 < t < 1.5 \cdot \tau_{trans}$ in small subintervals of length $L = \frac{1.5 \cdot \tau_{trans}}{50}$. For $t = 1.5 \cdot \tau_{trans}$ the fission width has already reached its stationary value. If we are inside a certain step $n$, we define the value of the fission-decay width as the quantity

$$\Gamma_{f,i}^n = \frac{\Gamma_f^n(t_{tot} + i \cdot L) + \Gamma_f^n(t_{tot} + (i+1) \cdot L)}{2} \quad (A1.11)$$

with $i = 0$ at the beginning of each step and $t_{tot} = t_{sum}^n + t_0$, where $t_0$ accounts for the zero point motion (see refs. [6, 7]). This value of the fission decay width is used to evaluate the decay probability inside this small subinterval $i$

$$P_{decay}^i = 1 - \exp(-\frac{L}{\tau^i}) \quad (A1.12)$$

with

$$\frac{1}{\tau^i} = \frac{\Gamma_p^n}{\hbar} + \frac{\Gamma_{f,i}^n}{\hbar} \quad (A1.13)$$

By means of a Monte-Carlo selection, we establish whether the nucleus decays inside this time subinterval $i$ or not. If no decay takes place, we evaluate again expression (A1.11) for the next subinterval $i+1$ and so on, until the nucleus decays. The decay channel is then determined by a further Monte-Carlo selection with the weights $\Gamma_p^n$ and $\Gamma_{f,i}^n$.

As shown in ref. [6], in this case we have an additional dependence of $\Gamma_f(t)$ on the deformation at the fission barrier $x_b$, on the reduced mass $\mu$ and on the frequency of the system at the ground state $\omega_g$. Considering that

$$\mu\omega_g^2 = K_g \qquad (A1.14)$$

where $K_g$ is the stiffness of the parabolic nuclear potential, the implementation of this description in the evaporation code requires determining the saddle-point deformation $x_b$ and the stiffness $K_g$ of the different nuclei that are produced during the deexcitation cascade. For the saddle point deformation $x_b$ we use the expression taken from [55]

$$x_b = \frac{7}{3}y - \frac{938}{765}y^2 + 9.499768 y^3 - 8.050944 y^4 \qquad (A1.15)$$

where $y = 1-\alpha$ and $\alpha$ is the fissility parameter. For the stiffness $K_g$, we use the liquid-drop-model predictions of reference [56]

$$K_g = 2 \cdot \left[ 7.1776 \left(1 - 1.7826 \left(\frac{A-2Z}{A}\right)^2\right) A^{2/3} - 0.1464 \frac{Z^2}{A^{1/3}} \right] \text{MeV} \qquad (A1.16)$$

If no decay occurs until the time limit $t = 1.5 \cdot \tau_{trans}$ is reached, the decay channel is determined at random according to the partial decay widths, which do not vary with time any more, just as in a conventional evaporation code.

For example, in reference [57] several analytical expressions for the probabilities of the diverse decay channels are given considering the step function and the exponential-like in-growth function to describe the form of $\Gamma_f(t)$. However, this is done under the simplifying assumption that the total decay widths are unchanged from one step to the next in the decay cascade. In reference [58], the exponential-like in-growth function is used, but the total decay time at each step of the cascade is considered to be ruled uniquely by the partial neutron-decay life time at that step, while the influence on the life time introduced by the other possible decay channels is not considered.

In view of the validity of this new approximation for $\Gamma_f(t)$ in the over-damped as well as in the under-damped regime, see ref. [6], we believe there is a clear improvement, and that this proposed description of the time-dependent fission-decay width may substitute less realistic formulations in nuclear-model codes, leading to more reliable conclusions on nuclear dissipation.